\documentclass[twocolumn]{aastex62}

\usepackage{amsmath,amssymb}
\usepackage{epstopdf}
\usepackage{ulem}

\submitjournal{ApJ}

\shorttitle{Electron SSA at SNR shocks}
\shortauthors{Bohdan et al.}

\newcommand{\rev}{\textcolor{black}}
\newcommand{\revrev}{\textcolor{black}}

\newcommand{\thbn}{\Theta_{\rm Bn}}
\newcommand{\degree}{^{\rm o}}
\newcommand{\ompe}{\omega_\mathrm{pe}}

\newcommand{\omce}{\Omega_\mathrm{e}}

\newcommand{\ms}{M_\mathrm{s}}
\newcommand{\ma}{M_\mathrm{A}}
\newcommand{\mi}{m_\mathrm{i}}
\newcommand{\me}{m_\mathrm{e}}
\newcommand{\lse}{\lambda_\mathrm{se}}
\newcommand{\lsi}{\lambda_\mathrm{si}}
\newcommand{\vsh}{v_\mathrm{sh}}

\begin{document}

\title{Kinetic simulations of nonrelativistic perpendicular shocks of young supernova remnants. I. Electron shock-surfing acceleration.}

\correspondingauthor{Artem Bohdan}
\email{artem.bohdan@desy.de}

\author[0000-0002-5680-0766]{Artem Bohdan}
\affil{DESY, 15738 Zeuthen, Germany}

\author{Jacek Niemiec}
\affil{Institute of Nuclear Physics Polish Academy of Sciences, PL-31342 Krakow, Poland}

\author{Martin Pohl}
\affil{DESY, 15738 Zeuthen, Germany}
\affil{Institute of Physics and Astronomy, University of Potsdam, 14476 Potsdam, Germany}

\author{Yosuke Matsumoto}
\affil{Department of Physics, Chiba University, 1-33 Yayoi-cho, Inage-ku, Chiba 263-8522, Japan}

\author{Takanobu Amano}
\affil{Department of Earth and Planetary Science, the University of Tokyo, 7-3-1 Hongo, Bunkyo-ku, Tokyo 113-0033, Japan}

\author{Masahiro Hoshino}
\affil{Department of Earth and Planetary Science, the University of Tokyo, 7-3-1 Hongo, Bunkyo-ku, Tokyo 113-0033, Japan}

\begin{abstract}

Electron injection at high Mach-number nonrelativistic perpendicular shocks is studied here for parameters that are applicable to young SNR shocks. Using high-resolution large-scale two-dimensional fully kinetic particle-in-cell (PIC) simulations and tracing individual particles we in detail analyze the shock surfing acceleration (SSA) of electrons at the leading edge of the shock foot. The central question is to what degree the process can be captured in 2D3V simulations. We find that the energy gain in SSA always arises from the electrostatic field of a Buneman wave. Electron energization is more efficient in the out-of-plane orientation of the large-scale magnetic field because both the phase speed and the amplitude of the waves are higher than for the in-plane scenario. Also, a larger number of electrons is trapped by the waves compared to the in-plane configuration. We conclude that significant modifications of the simulation parameters are needed to reach the same level of SSA efficiency as in simulations with out-of-plane magnetic field or 3D simulations.

\end{abstract}


\section{Introduction}\label{introduction}

The current paradigm of cosmic-ray (CR) origin assumes that the most part of galactic CR population is produced at nonrelativistic forward shocks of supernova remnants (SNRs).
The main acceleration mechanism considered at shocks is diffusive shock acceleration (DSA), a~first-order Fermi process \citep[e.g.,][]{1977ICRC...11..132A,1983RPPh...46..973D,1987PhR...154....1B}.
Astronomical observations give strong support to this paradigm. In particular, detection of broadband nonthermal emission from SNRs, extending in some objects to TeV-range gamma rays, proves the presence of ultrarelativistic particles in these sources, though for most SNRs it is still unclear which parent particle populations (protons or electrons) generate dominant high-energy emission \citep{2013APh....43...71A}. 

Acceleration of particles through DSA comes from multiple interactions with the shock front, while they bounce between the shock upstream and downstream plasmas. Particle confinement to the shock vicinity is provided by elastic scattering off magnetohydrodynamic (MHD) turbulence that renders diffusive particle motions.       
The critical ingredient and the main unsolved problem in the DSA theory is the particle injection.
CRs undergoing DSA have Larmor radii much larger than the internal shock transition width, that is commensurate with the gyroradius of the incoming protons (with shock speed $\vsh$). CRs thus see the shock as a sharp discontinuity in the plasma flow. To be fed into the acceleration process particles need therefore to be extracted from the thermal pool and pre-accelerated. 
Since protons have a larger initial momentum and can be easily scattered either by MHD waves embedded in the ambient plasma or by self-generated turbulence, their injection is relatively easy to account for. The problem is more severe for electrons, because of their smaller mass and consequently smaller gyroradii and inertial lengths, compared to protons, and is known as the electron injection problem.

\begin{figure}[t!]
\centering
\includegraphics[width=\linewidth]{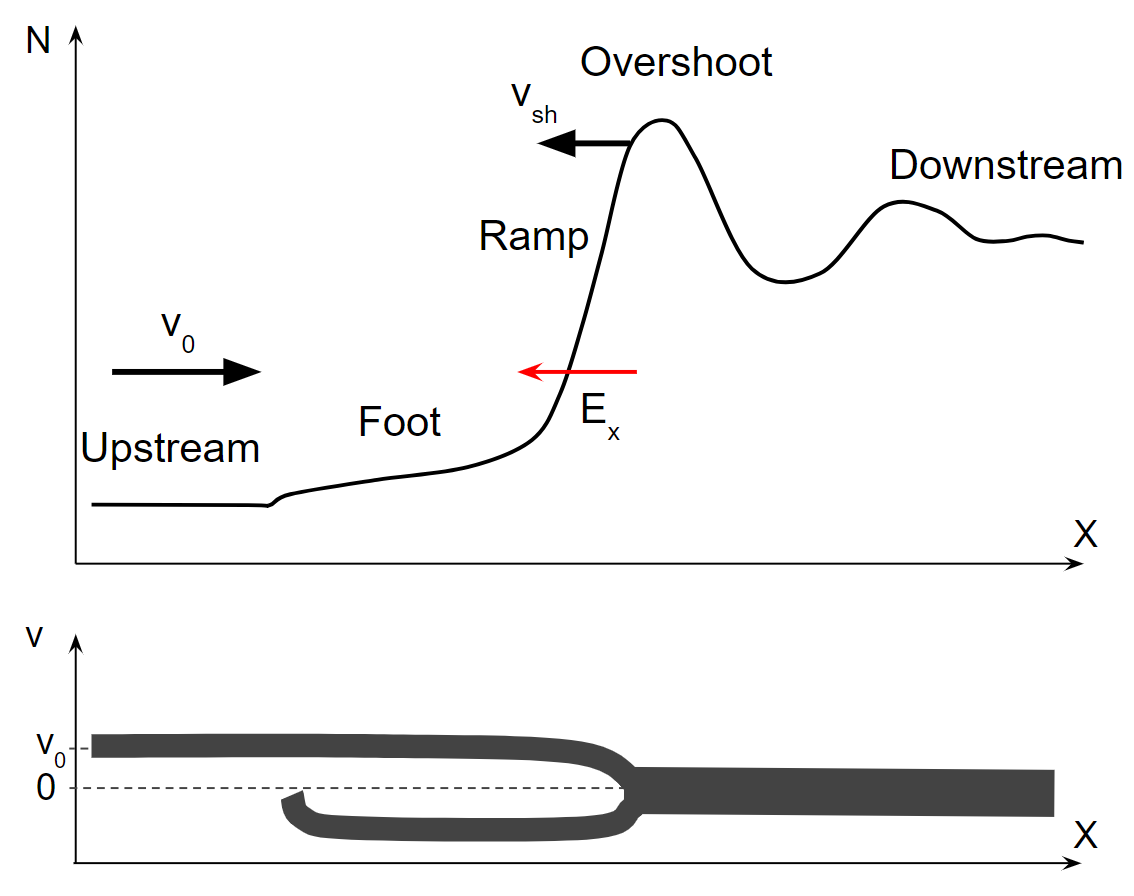}
\caption{Perpendicular shock structure. Top panel is the particle number density profile. The shock transition consists of a foot, a ramp, an overshoot and the downstream region. $E_x$ is the shock potential. $v_0$ and $v_{sh}$ are the upstream and the shock velocities. Bottom panel is the x-component of ion phase-space distribution.}
\label{shock_schema}
\end{figure}

Here we study electron injection at young SNR shock waves using particle-in-cell (PIC) numerical simulations that provide a fully self-consistent treatment of the electron scales. Such shocks are characterized by high sonic, $\ms$, and Alfv\'enic, $\ma$, Mach numbers.
Present observational data do not give clear constraints on the large-scale magnetic-field configuration in portions of SNR shocks from which strong nonthermal emission is detected. Radio polarimetry are notoriously difficult to interpret \citep[e.g.,][]{2009ApJ...696.1864S}.
Different approaches of data modeling for the same source can suggest the presence of quasi-perpendicular fields \citep{2009MNRAS.393.1034P,2010MNRAS.408..430S,2016A&A...587A.148W} or the opposite, quasi-parallel configurations (\citealp{2004A&A...425..121R}; \citealp{2011A&A...531A.129B,2015MNRAS.449...88S}).
As in our recent studies \citep{2012ApJ...755..109M,2013PhRvL.111u5003M,2015Sci...347..974M,2016ApJ...820...62W,2017ApJ...847...71B}, in this work we examine perpendicular shocks as the most simple form of a quasi-perpendicular magnetic-field configuration.
The physics of such shocks is governed by reflection of ions at the shock caused by shock potential (Fig.~\ref{shock_schema}), the interaction of which with the incoming plasma excites a variety of instabilities upstream of the shock. The most important instabilities in the regime of high Mach numbers are the electrostatic two-stream Buneman instability at the leading edge of the foot, resulting from the interaction between cold incoming electrons and reflected ions \citep{Buneman1958}, and the Weibel instability in the shock foot driven by the interaction of the incoming and reflected ions \citep{2010ApJ...721..828K,2012ApJ...759...73N,2015Sci...347..974M,2016ApJ...820...62W}.

The Buneman instability can mediate the generation of supra-thermal electrons via shock surfing acceleration (SSA). In a 1D picture the Buneman instability produces strong, coherent electrostatic waves that capture electrons and let them be accelerated by the convective electric field \citep{2002ApJ...572..880H}, thus providing for efficient electron injection. A number of 2D simulations of perpendicular shocks \citep{2009ApJ...690..244A,2012ApJ...755..109M,2013PhRvL.111u5003M,2016ApJ...820...62W} demonstrated that the length of the potential wells is limited to about the ion inertial length. Electrons can thus escape from the trapping region and re-enter it from the downstream or the upstream side to experience multiple surfing-acceleration events \citep{2009ApJ...690..244A,2012ApJ...755..109M}. 

The Weibel instability generates strong magnetic fields with filamentary structure.
It was also recently shown with 2D simulations that spontaneous turbulent magnetic reconnection in the Weibel instability region canlead to electron acceleration \citep{2015Sci...347..974M}. Thin current sheets (magnetic filaments) 
become unstable and break up into chains of magnetic islands and X-points. Particles can be accelerated while interacting with these structures.

The spectrum of waves generated at the shock is usually at least two-dimensional. Which of the unstable modes appear in a 2D simulation strongly depends on the configuration of the mean magnetic field though, as modes may be artificially suppressed if their wave vector is not contained in the simulation plane. In \citet{2017ApJ...847...71B} we showed that the
Weibel instability is best reproduced with the in-plane setup, whereas the Buneman modes are considerably stronger and more coherent with a strictly out-of-plane orientation.
Suprathermal tails in the electron spectra are found for all simulated shocks, and the initial acceleration of electrons always occurs through the SSA process in the~Buneman wave region. However, the subsequent stages of injection strongly depend on the field configuration. For out-of-plane field adiabatic heating dominates the spectral evolution. For configurations with an in-plane magnetic-field component particles are non-adiabatically accelerated in interactions with turbulent magnetic structures in the shock, resembling a second-order Fermi process, and magnetic reconnection does also occur. The fraction of nonthermal electrons is an order of magnitude larger for the out-of-plane configuration than for other field orientations, mainly on account of a higher SSA efficiency.

The first 3D PIC simulation of a high-$\ma$ shock was recently presented by \cite{Matsumoto2017} for an oblique \emph{subluminal} configuration,  $c / \tan{\thbn}>\vsh$, where $\thbn$ is the angle of the large-scale magnetic field with respect to the shock normal, $\vsh$ is the shock velocity, and $c$ is the speed of light. Buneman waves and Weibel magnetic turbulence were found to coexist in the shock structure. Energetic electrons that initially experienced SSA underwent pitch-angle diffusion by interacting with magnetic turbulence in the shock foot and ramp. This provides confinement in the shock transition region during which particles gain energy by shock drift acceleration (SDA). The computational cost of 3D experiments is still too high to sample the range of plasma conditions that one may find in SNR shocks. Nevertheless, the 3D results indicate which parts of 3D shock physics can be reliably probed with 2D simulations. 

In this work we report on new large-scale 2D fully kinetic PIC simulations of nonrelativistic strictly perpendicular shocks in the regime of high Mach numbers, $\ma\gtrsim 20$ and $\ms\gtrsim 30$, as appropriate for forward shocks of young SNRs. The simulations are conducted in 2D3V configuration, i.e., we follow two spatial coordinates and all three components of the velocity and the electromagnetic fields. Numerical experiments are performed for both in-plane and out-of-plane configurations of the large-scale magnetic field. 
These simulations complement our previous investigations of 2D perpendicular shocks \citep[e.g.,][]{2012ApJ...755..109M,2013PhRvL.111u5003M,2015Sci...347..974M,2016ApJ...820...62W,2017ApJ...847...71B}. 
The aim of this work is to analyze in detail the initial energization via SSA in the Buneman-instability region. The successive acceleration in the shock foot and ramp on account of, e.g., inelastic scattering off the Weibel-instability turbulence is the subject of a~separate publication.

Conditions for efficient electron energization via SSA were first investigated by \cite{2012ApJ...755..109M}, supported with PIC simulations with out-of-plane magnetic-field configuration. The process occurs in low-temperature (low beta) plasmas, in which the Buneman instability can effectively grow. 
For efficient acceleration the electrostatic waves should also be strong enough to trap electrons and hold them during acceleration, which defines a minimum Alfv\'enic Mach number for a shock to be capable of producing relativistic electrons via SSA,
\begin{equation}
  \ma \geq (1+\alpha)  \left(\frac{m_\mathrm{i}}{m_\mathrm{e}}\right)^{\frac{2}{3}},
  \label{trapping}
\end{equation}
where $\alpha$ is the flux ratio of reflected to incoming ions and $\mi$ and $\me$ are the ion and the electron mass, respectively. In the presence of an in-plane magnetic field the motion of the reflected ions is not fully contained in the simulation grid and thus the corresponding component of the Buneman waves cannot be captured \citep{2017ApJ...847...71B}. To account for this effect we proposed a modified trapping condition:
\begin{equation}
 \ma \geq \sqrt\frac{2}{1+\sin^2\varphi}
 (1+\alpha)  \left(\frac{m_\mathrm{i}}{m_\mathrm{e}}\right)^{\frac{2}{3}},
\label{trappingnew}
\end{equation}
where $\varphi$ is the orientation angle of the large-scale perpendicular magnetic field with respect to the simulation plane, with $\varphi=0\degree$ representing the in-plane configuration (see Fig.~\ref{setup}). The earlier 2D simulations of \citet{2017ApJ...847...71B} all satisfied the trapping condition of Equation~\ref{trapping} and were performed for a single value of the reduced mass ratio, $\mi/\me=100$, and a small ($\beta_{\rm e}\ll 1$) or moderate ($\beta_{\rm e}=0.5$) plasma beta. Our present work augments this analysis with investigations of the trapping conditions of Equations~\ref{trapping} and~\ref{trappingnew} and SSA efficiency for different mass ratios in the range $\mi/\me=50-400$.
\citet{Matsumoto2017} demonstrated that the SSA process is well reproduced with 2D out-of-plane simulations, but processes in the shock ramp and overshoot are suppressed. On the other hand, the stochastic Fermi-like acceleration in the Weibel-instability-generated turbulence works similar as in 2D in-plane experiments. If the modified trapping condition \rev{would define the parameter range, for which we have} the same efficiency of electron pre-acceleration for an in-plane configuration as that observed in 2D simulations with out-of-plane magnetic field, it would be possible to reproduce realistic 3D physics with far cheaper 2D experiments with $\varphi=0\degree$. \rev{This is the main hypothesis under discussion here.}

The paper is organized as follows. We present a description of the simulation setup in Section~\ref{sec:setup}. The results are presented in Section~\ref{results}. Section~\ref{summary} contains the summary and discussion.


\section{Simulation Setup} \label{sec:setup}

\begin{figure}[htb]
\centering
\includegraphics[width=\linewidth]{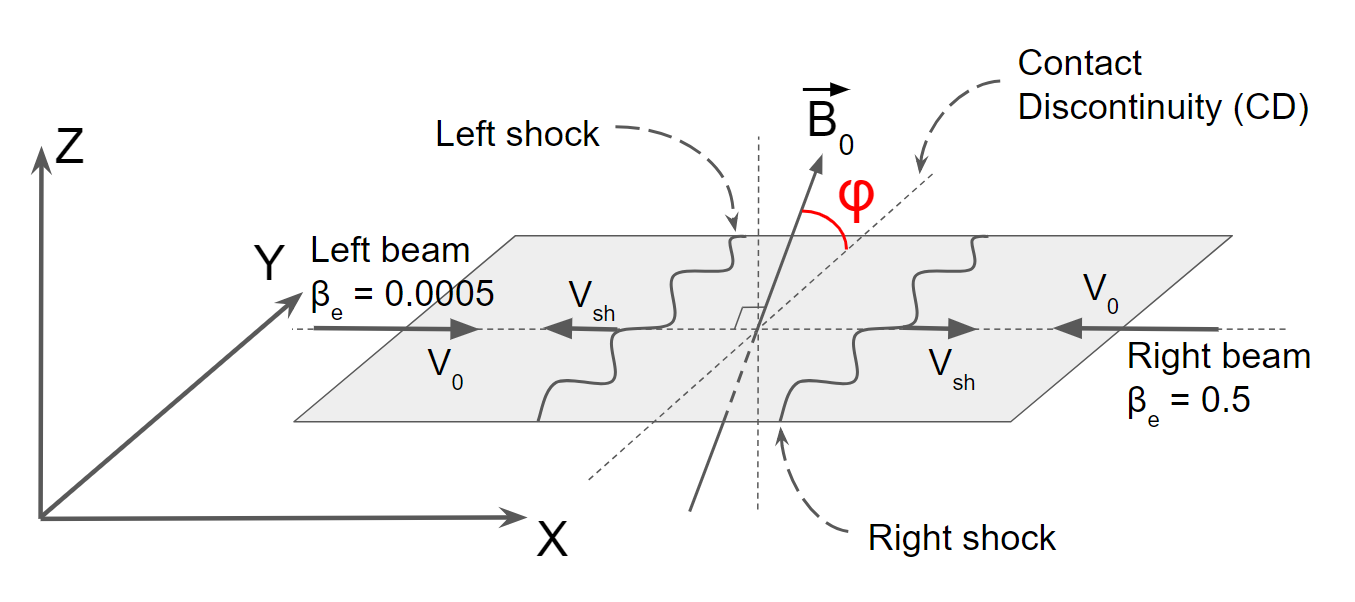}
\caption{Illustration of the simulation setup.} 
\label{setup}
\end{figure}

   \begin{table*}[!t]
      \caption{Simulation Parameters}
         \label{table-param}
     $$ 
\begin{array}{p{0.07\linewidth}rcrccrcccccr}
\hline
\hline
\noalign{\smallskip}
Runs   & \varphi  & L_y (\lambda_{\rm si}) & \mi/\me & \omega_{\rm pe}/\Omega_{\rm e} &  \ma & \multicolumn{2}{c}{\ms} & \multicolumn{2}{c}{\beta_{\rm e}} & Eq.~\ref{trapping} & \multicolumn{2}{c}{Eq.~\ref{trappingnew}}\\
 & & & & & & ^*1 & ^*2 & ^*1 & ^*2 & \alpha=0.2  & \alpha=0.2 & (0.5) \\
\noalign{\smallskip}
\hline
\noalign{\smallskip}
A1, A2  & 0^o & 10.9 & 50  & 12 &  22.6  & 949 & 30 & 5 \cdot 10^{-4} & 0.5 & 16  & 22.4 & (28) \\
B1, B2  & 0^o &  24  & 100 & 12 &  31.8  & 1342 & 42.4 & 5 \cdot 10^{-4} & 0.5 & 26  &  36 & (46)\\
C1, C2  & 0^o &  12  & 100 & 17.3 &  46  & 1941 & 61.4 & 5 \cdot 10^{-4} & 0.5 & 26  &  36 & (46)\\
D1, D2  & 0^o &  11.9  & 200 & 8.5 &  32  & 1342 & 42 & 5 \cdot 10^{-4} & 0.5 & 41  &  58 & (72)\\
E1, E2  & 0^o &  11.9  & 200 & 12 &  44.9  & 1898 & 60 & 5 \cdot 10^{-4} & 0.5 & 41  &  58 & (72) \\
F1, F2  & 0^o &  8.2  & 400 & 12 & 68.7  & 2904 & 91.8 & 5 \cdot 10^{-4} & 0.5 & 65  &  92 & (115)\\
\noalign{\smallskip}
\hline
\noalign{\smallskip}
G1, G2  & 90^o & 12  & 100 & 12 & 35.5  & 1369 & 43.3 & 5 \cdot 10^{-4} & 0.5 & 26  &  36 & (46)\\
\noalign{\smallskip}
\hline
\end{array}
     $$ 
\tablecomments{Parameters of simulation runs described in this paper. Listed are: the orientation of the uniform perpendicular magnetic field with respect to the 2D simulation plane, $\varphi$, the transverse size of the computational box, $L_y$, in units of the ion skin depth, $\lambda_{\rm si}$, the ion-to-electron mass ratio $\mi/\me$, the plasma magnetization, $\omega_{\rm pe}/\Omega_{\rm e}$, and Alfv\'enic  and sonic Mach numbers, $\ma$ and $\ms$, the latter separately for the \emph{left} (runs *1) and the \emph{right} (runs *2) shock. We also list the electron plasma beta, $\beta_{\rm e}$, for each simulated shock and the critical Alfv\'enic Mach number (Eq.~\ref{trapping}) for $\alpha=0.2$, as well as the modified trapping condition (Eq.~\ref{trappingnew}) calculated for $\alpha=0.2$ and  $\alpha=0.5$ (in brackets).
All runs use the electron skin depth of $\lambda_{\rm se}=20\Delta$.} 
   \end{table*}

The simulation setup adopted in this work is the same as that used in \cite{2017ApJ...847...71B} and illustrated in Figure~\ref{setup}. As a result of the collision of two counter-streaming electron-ion plasma beams, two shocks are formed that propagate in opposite directions and are separated by a contact discontinuity (CD). The plasma flow is set along the $x$-direction in the $xy$ plane. Plasma particles are continuously injected at both sides of the simulation box with velocities $\boldsymbol{v}_{\rm L}=v_{\rm L}\hat{\boldsymbol{x}}$ and 
$\boldsymbol{v}_{\rm R}=v_{\rm R}\hat{\boldsymbol{x}}$, where the indices L~and R refer, respectively, to the \emph{left} and \emph{right} sides of the simulation box. As the two shocks move away from the CD in the left and the right plasma, we refer to them as to the \emph{left} and the \emph{right} shocks, respectively.
The two plasma streams carry a homogeneous magnetic field, $\boldsymbol{B}_{\rm 0}$, that is perpendicular to the shock normal and lies in the $yz$ plane. The magnetic field thus forms an angle $\varphi$ with the $y$-axis. Initialized with the flow is a motional electric field $\boldsymbol{E}_{\rm 0}=-\boldsymbol{v}\times\boldsymbol{B}_{\rm 0}$, with $\boldsymbol{v}=\boldsymbol{v}_{\rm L}$ or $\boldsymbol{v}=\boldsymbol{v}_{\rm R}$, respectively, for the left and the right beam.
We assume that the beams move with equal absolute velocities, $v_{\rm L}=v_{\rm R}=0.2c$, and that the 
magnetic field strength in both plasmas is equal, $\boldsymbol{B}_{\rm 0L}=\boldsymbol{B}_{\rm 0R}$. The motional electric field thus has equal strength and opposing signs in the two slabs. We use the method of \citet{2016ApJ...820...62W} to suppress the artificial electromagnetic transient that results from the initial strong electric-field gradient between the two plasma slabs.

We collide plasma beams of equal density but different temperatures, thus studying two different shocks in one simulation. The temperature ratio between the two beams is $1000$, so that the \emph{sonic} Mach numbers, $M_{\rm s}$, of the two shocks differ by a factor of $\sqrt{1000} \simeq 30$. In terms of the \emph{electron} plasma beta (the ratio of the electron plasma pressure to the magnetic pressure) the left beam has $\beta_{\rm e,L}=5 \cdot 10^{-4}$ and the right beam $\beta_{\rm e,R}=0.5$.  
This choice of plasma beta facilitates a direct comparison with our earlier work \citep{2017ApJ...847...71B} and also with results of previous 2D simulations of perpendicular shocks \citep{2012ApJ...755..109M,2013PhRvL.111u5003M} and a recent 3D simulation of a quasi-perpendicular shock \citep{Matsumoto2017}, in which $\beta_{\rm e,R}=0.5$ is assumed.
Note that our system is approximately in ram-pressure balance, and consequently the simulation frame is also the \emph{downstream} rest frame of the two shocks.

\begin{figure}[b!]
\centering
\includegraphics[width=\linewidth]{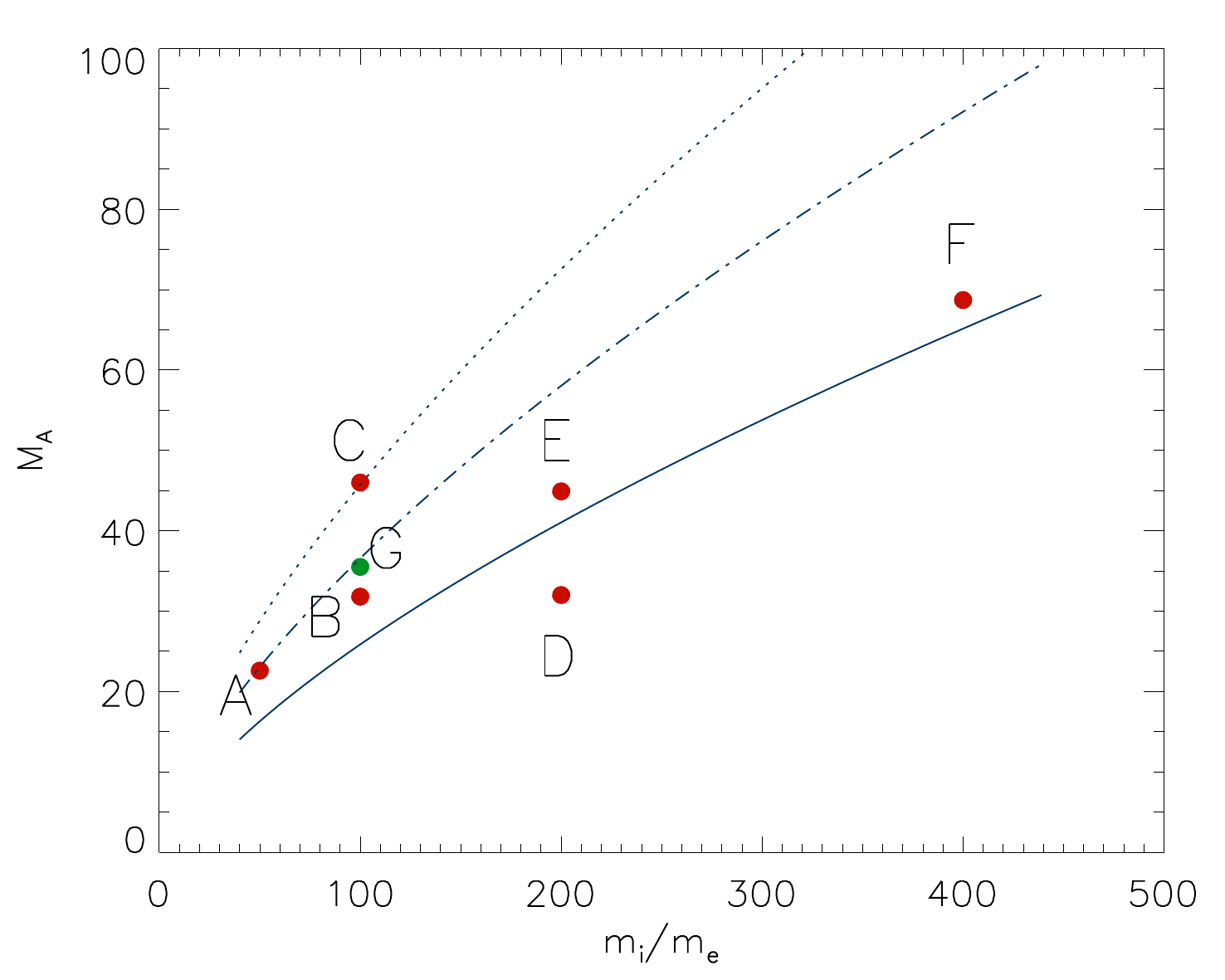}
\caption{The Alfv\'enic Mach numbers and mass ratios of the simulation runs. Runs A--F with in-plane magnetic field configuration are depicted with red dots. Run G with the out-of-plane field is marked with a green dot. The blue solid line shows the scaling given by the trapping condition of Eq.~\ref{trapping}, calculated for $\alpha=0.2$. The \rev{blue} dash-dotted and dotted lines show the modified trapping condition (Eq.~\ref{trappingnew}) for $\alpha=0.2$ and $\alpha=0.5$, respectively.}  
\label{parameters}
\end{figure}

The parameters of the simulation runs described in this paper are listed in Table~\ref{table-param}.
We have performed seven large-scale numerical experiments (runs A--G), that feature in total fourteen simulated shocks. Here we refer to each of these shock cases as to a separate simulation run, and tag the shocks in the left plasma ($\beta_{\rm e,L}=5 \cdot 10^{-4}$) with *1, and the right shocks with *2 ($\beta_{\rm e,R}=0.5$). Simulation runs A--F assume the in-plane magnetic field configuration, $\varphi=0^o$, and run G uses the out-of-plane magnetic field orientation, $\varphi=90^o$.
We do not consider simulations with $\varphi=45^o$, because the shock structure and the acceleration mechanisms observed in this case are almost identical to those in runs with the in-plane field configuration \citep{2017ApJ...847...71B}.
The runs with the in-plane magnetic field cover a wide range of ion-to-electron mass ratios and Alfv\'enic Mach numbers, as illustrated in Figure~\ref{parameters}, which permits an investigation of the influence of these parameters on the electron acceleration efficiency and to scale our results to the realistic ion-to-electron mass ratio. Note, that some aspects of the shock physics in runs B and G have been already discussed in our previous paper \citep[cf. runs A and C in][]{2017ApJ...847...71B}.

The derived shock properties are also listed in Table~\ref{table-param}.
The Alfv\'en velocity is defined as $v_{\rm A}=B_{\rm 0}/\sqrt{\mu_{\rm 0}(N_e\me+N_i\mi)}$, where $\mu_{\rm 0}$ is the vacuum permeability, $N_i$ and $N_e$ are the ion and the electron number densities, and $B_0$ is the far-upstream magnetic-field strength. The sound speed reads $c_{\rm s}=(\Gamma k_BT_{\rm i}/\mi)^{1/2}$, where $k_B$ is the Boltzmann constant, $\Gamma$ is a nonrelativistic adiabatic index, and $T_{\rm i}$ is the ion temperature.
The Alfv\'enic, $\ma=\vsh/v_{\rm A}$, and sonic, $\ms=\vsh/c_{\rm s}$, Mach numbers of the shocks in Table~\ref{table-param} are given in the conventional \emph{upstream} reference frame. 
As the in-plane and the out-of-plane magnetic field lead to a different number of degrees of freedom, the adiabatic indices are different with $\Gamma=5/3$ and $\Gamma=2$, respectively for $\varphi=0^o$ and $\varphi=90^o$. Thus the resulting expected shock speeds take values $\vsh=0.263c$ for runs A--F and $\vsh=0.294c$ for runs G. In the simulation frame the speeds are smaller by the shock compression ratio.

To investigate the role of SSA in electron pre-acceleration, we adjust the magnetic-field strength, $B_0$, to establish Alfv\'enic Mach numbers that test the trapping conditions defined by Equations~\ref{trapping} and~\ref{trappingnew}. A comparison of the Alfv\'enic Mach numbers and the mass ratio of all runs with trapping limits is offered in Figure~\ref{parameters}.  Nevertheless, we always consider weakly magnetized plasmas with the ratio of the electron plasma frequency, $\omega_{\rm pe}=\sqrt{e^2N_e/\epsilon_0\me}$, to the electron gyrofrequency, $\Omega_{\rm e}=eB_0/\me$, in the range $\omega_{\rm pe}/\Omega_{\rm e}=8.5-17.3$. Here, $e$ is the electron charge, and $\epsilon_0$ is the vacuum permittivity. To keep the plasma beta constant we adjust the plasma temperatures and hence the sound speeds and resulting sonic Mach numbers (see Table~\ref{table-param}).

In this work we want to verify several hypotheses.   
The first is the scaling of the SSA efficiency with the ion-to-electron mass ratio for shocks that fulfill the trapping condition of Equation~\ref{trapping}, here applied to the in-plane magnetic field configurations. Runs A, B, E, and F define the set of simulations conducted for $\mi/\me=50, 100, 200$, and $400$, respectively.

The second objective is the modified trapping condition of Equation~\ref{trappingnew}. We test this condition by conducting simulation runs~C, which satisfy Equation~\ref{trappingnew} for $\alpha\leq 0.5$. The question to be addressed is whether 2D simulations with in-plane magnetic field configuration can reproduce the SSA efficiency observed in 2D runs with the same $\mi/\me$ and the out-of-plane fields, here marked as runs~G.    

The third set of simulations consists of runs D and E, performed for the same mass ratio $\mi/\me=200$. The Alfv\'enic Mach number in run D clearly violates Equation~\ref{trapping}, and so we expect a very low intensity of Buneman waves. Nevertheless, particle acceleration can still occur in the shock foot and ramp, whose structure is defined by the magnetic filaments, and we are interested in the nonthermal electron population that forms in the absence of SSA. Note, that cross-comparison of runs B and D, and C and E can yield the mass-ratio dependence for shocks having the same Alfv\'enic Mach numbers.

The electron skin depth in the upstream plasma is common for all runs and equals $\lse=20\Delta$, where $\Delta$ is the size of grid cells. The ion skin depth, $\lsi=\sqrt{\mi/\me}\lse$, is used here as the unit of length. 
The time scale and all temporal dependencies are given in terms of the upstream ion Larmor frequency, $\Omega_{\rm i}$, where $\Omega_{\rm i}=eB_0/\mi$. The simulation time is typically $t=(6-8)\Omega_i^{-1}$, which is enough to cover at least a few shock self-reformation cycles \citep[see][]{2017ApJ...847...71B}. The time-step we use is $\delta t=1/40\,\omega_{\rm pe}^{-1}$.

The two plasma beams injected at sides of the simulation box are composed of an equal number of ions and electrons, $N_{\rm ppc}=20$. Electron and ion plasma pairs are initialized at the same locations to ensure the initial charge-neutrality of the system. There is no escape of particles from the computational box, and we use injection layers receding from the CD as in \citet{2017ApJ...847...71B}, which helps alleviating numerical grid-Cerenkov effects and saves computational resources.   
The simulation box expands in $x$-direction during the run. The final size of a simulation box can reach $L_x\approx280\lambda_{\rm si}$.
The transverse size of the simulation box, $L_y=(8.2-24) \lambda_{\rm si}$, is large enough to cover several of the magnetic filaments, that are typically separated by $\sim\lsi$, and at the same time limits the computational expense that grows quadratically with $\mi/\me$. The largest simulation box of size $L_x \times L_y=(3264 \times 96000)\Delta$ is used in run F with $\mi/\me=400$. Open boundary conditions are imposed in the $x$-direction and periodic boundaries are applied in the $y$-direction. 

The numerical code we use is a 2D3V-adapted and modified version of the relativistic electromagnetic PIC code TRISTAN \citep{tristan} with MPI-based parallelization \citep{niemiec_2008,2016ApJ...820...62W} and the option to trace individual particles.


\section{Results} \label{results}

In Section~\ref{buneman-inst} we describe the structure of the Buneman wave modes in all simulations and also summarize the findings of \cite{2017ApJ...847...71B}. Then we discuss the electron acceleration efficiency through SSA in Section~\ref{ele-acc-buneman}.     


\subsection{The Buneman Instability} \label{buneman-inst}

\begin{figure*}[!t]
\centering
\includegraphics[width=0.98\linewidth]{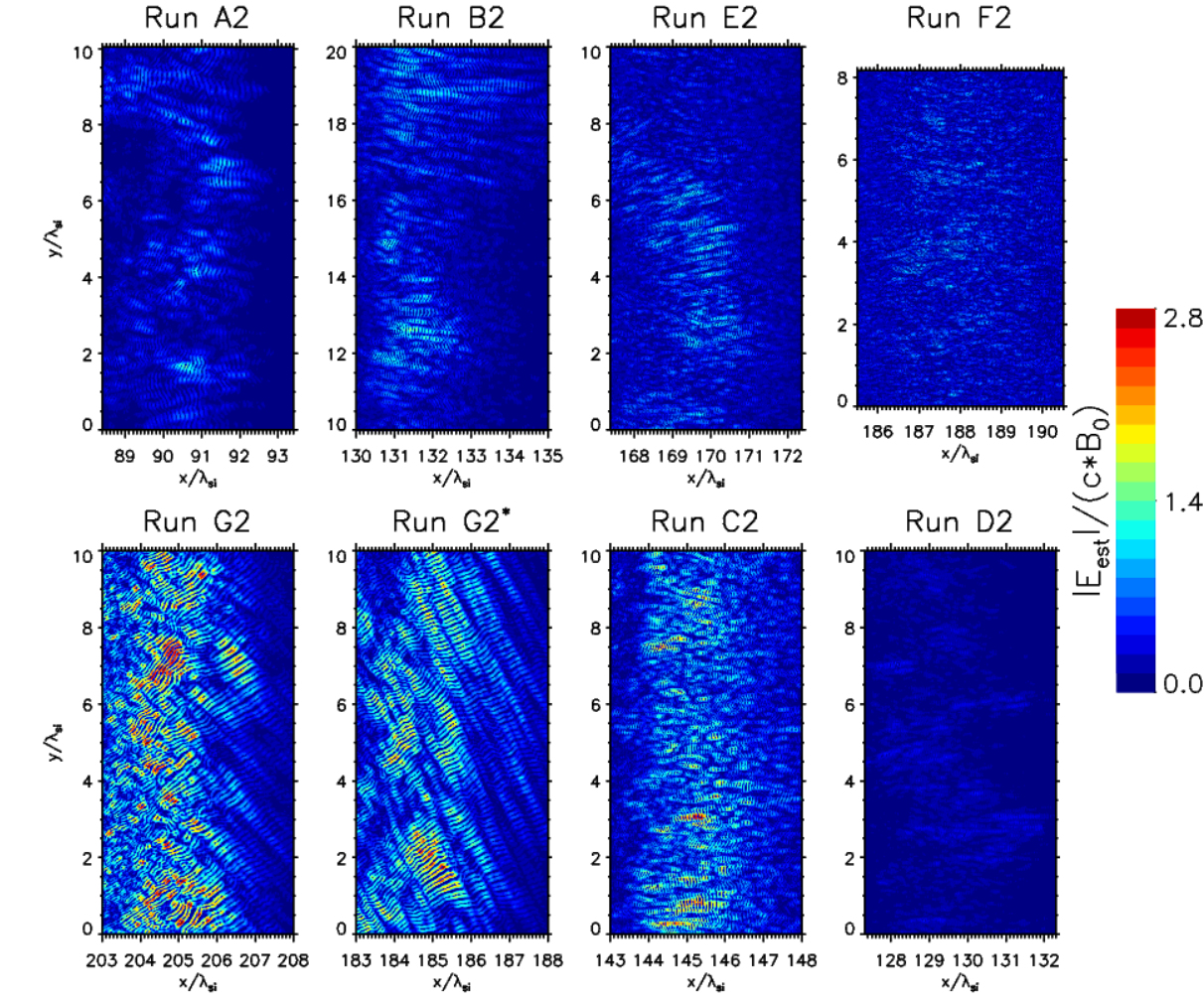}
\caption{Dimensionless electrostatic field amplitudes in selected regions of the shock foot with the most intense Buneman waves for runs~2. The map marked as run G2* is chosen at time moment when the average field strength is the same as in run C2. }
\label{Eest_plots}
\end{figure*}

Figure~\ref{Eest_plots} presents the maps of the electrostatic field amplitude in the foot of the \emph{right} shocks (runs A2-G2, see Table~\ref{table-param}), that propagate in moderate-temperature plasmas with $\beta_{\rm e}=0.5$. Only portions of the simulation boxes are shown to facilitate one-to-one comparison between the runs. Run G2* is run G2 at a different phase of shock reformation. The electrostatic fields are calculated as $|E_\mathrm{ES}|=|-\nabla\phi|$, where $\phi$ is the electric potential, that is derived directly from the charge distribution. The maps are plotted for simulation times, at which the cyclic shock self-reformation allows the strongest Buneman modes. Note, that the maps for runs B2 and G2 can be compared with Figures 6a3 and 6c3, respectively, in \cite{2017ApJ...847...71B}, in which results for runs B1 and G1 are presented (marked as runs A1 and C1, respectively). 

The properties of the Buneman instability discussed in \cite{2017ApJ...847...71B} can be readily observed in Figure~\ref{Eest_plots}. The wave vectors are approximately parallel to the shock normal for the in-plane configurations (runs A2-F2) and oblique for out-of-plane magnetic field (run G2). This reflects the motion of shock-reflected ions: for $\varphi=0\degree$ the ions are confined to the $xz$-plane whereas for $\varphi=90\degree$ they stream in the simulation plane. The Buneman wave region shows a patchy structure for the in-plane field configurations, that can be linked to clumps in the overshoot produced by merging magnetic filaments. 
In total, the Buneman waves occupy a much smaller region than for the out-of-plane configuration, for which the waves are coherent and more intense. 

The phase velocity of the Buneman modes matches the relative speed between shock-reflected ions and incoming electrons of the upstream plasma. Since for $\varphi=0^o$ part of the ion motion is outside of the simulation grid, the wavelengths of the Buneman waves are smaller ($\lambda\approx 1.9\lse$) than for out-of-plane field, for which $\lambda\approx 3.3\lse$. Note, that Figure~\ref{Eest_plots} shows $\vert E\vert$ and hence the wavelength is twice the separation of wave fronts, here provided in units of the ion skin depth. The surface area of the Buneman wave region for shocks in moderate-temperature plasma is 20\%-30\% larger than at the corresponding low-$\beta$ shocks, but the intensity of the waves is 20\%-50\% smaller \citep[compare Fig. 6 in][]{2017ApJ...847...71B}.

For the high-$\beta$ systems presented in Figure~\ref{Eest_plots}, Table~\ref{table-Buneman} lists peak amplitude of Buneman waves and the fraction of pre-accelerated electrons. 
The runs A2, B2, E2 and F2 satisfy the trapping condition of Equation~\ref{trapping} (see Fig.~\ref{parameters}), and both the peak and average strength of the electrostatic field are similar. Small differences between them arise from shock reformation. We conclude that irrespective of the mass ratio, the physical conditions at shocks with $\ma$ satisfying Equation~\ref{trapping} are similar. However, the electrostatic force is weaker in average than the Lorentz force on a $\gamma\gtrsim 2$ electron ($|E_\mathrm{ES}|/(cB_0) <1$). 

\begin{table}[!th]
      \caption{Dimensionless peak amplitude of Buneman waves and fraction of pre-accelerated electrons}
         \label{table-Buneman}
\centering
\begin{tabular}{lccc}
\hline
\hline
\noalign{\smallskip}
Run   & $\max(|E_\mathrm{ES}|/(cB_0))$ & $N_{\rm e,BI}/N_{\rm e,tot} (\%)$ \\
\noalign{\smallskip}
\hline
\noalign{\smallskip}
A2 &  1.1   &   0.43  \\
B2 &  1.3   &   0.46  \\
C2 &  2.3   &   0.6   \\
D2 &  0.4   &   0.34  \\
E2 &  1.3   &   0.49  \\
F2 &  1.1   &   0.44  \\
G2 &  2.7   &   6.8   \\
G2*&  2.3   &   2.7   \\
\noalign{\smallskip}
\hline
\end{tabular}
\smallskip
\tablecomments{For $\beta_{\rm e}=0.5$ shocks we list the normalized peak amplitude, $\max(|E_\mathrm{ES}|/(cB_0))$, of the electrostatic waves, calculated as mean $|E_\mathrm{ES}|/(cB_0)$ for the 100 simulation cells with the highest $|E_\mathrm{ES}|/(cB_0)$, and the fraction of electrons pre-accelerated to $(\gamma-1)>0.1$.}
   \end{table}

Considerably larger electrostatic field amplitudes, reaching $|E_\mathrm{ES}|/(cB_0)\sim\! 2.3$, can be observed for run C2. Here, the Alfv\'en Mach number of the shock, $\ma=46$, is much larger than the minimum $\ma$ defined by Equation~\ref{trapping} and also satisfies the modified trapping condition of Equation~\ref{trappingnew}, that for the measured $\alpha\simeq 0.32$ gives the minimum $\ma\simeq 40.2$. The field intensity in run C2 is about a factor of 2 larger than in both run B2 with the same mass ratio, $\mi/\me=100$, and  run E2 with mass ratio $\mi/\me=200$ but similar Alfv\'en Mach number, $\ma\simeq 45$. This shows that the strength of the electrostatic modes is driven by the value of the Alfv\'enic Mach number in relation to the trapping condition (Eq.~\ref{trapping}). The absolute value of $\ma$ is not important, as in run D2 we see Buneman waves with amplitudes a factor of 3 lower than those in run B2 with the same Alfv\'enic Mach number. Essentially all observed wave intensities are slightly weaker than the saturation level estimated by \citet{1980PhRvL..44.1404I}.

The modified trapping condition (Eq.~\ref{trappingnew}) was expected to compensate for the effect of the field configuration. Shocks with sufficiently large $\ma$ should then reproduce similar Buneman wave intensities in 2D in-plane magnetic field configurations than in simulations with   
the out-of-plane fields.
However, the electrostatic field in run C2 is weaker by 20\% than that in run G2 with $\varphi=90^o$. At a different phase of shock reformation run G2, now called G2*, has the same electric-field amplitude as C2, but four times the number of pre-accelerated electrons. This discrepancy might arise from  Equation~\ref{trappingnew} only compensating for the neglect of the $z$-motion of ions. In out-of-plane simulation we observe 
that the relative speed between electrons and reflected ions can reach $\sim0.6c$, because of acceleration in upstream electric field, which is a factor of $\sim1.5$ larger then the value assumed in the derivation of the trapping condition \citep[see][]{2012ApJ...755..109M}, while in in-plane case the acceleration is in $z$-direction. 
It may be that we need to also account for this effect by adding a factor of 1.5 to the modified trapping condition,
\begin{equation}
 \ma \geq 1.5 \sqrt\frac{2}{1+\sin^2\varphi}
 (1+\alpha)  \left(\frac{m_\mathrm{i}}{m_\mathrm{e}}\right)^{\frac{2}{3}}.
\label{trappingnew2}
\end{equation}
This equation gives $\ma\simeq 60.3$ for the minimum Alfv\'en Mach number, with which the amplitudes of the Buneman waves observed at shocks with $\ma = 35.5$ in 2D simulations with out-of-plane magnetic fields could be reproduced in runs applying $\varphi=0^o$ field configuration. This value is much larger than any of the Mach numbers studied here for $\mi/\me=100$, and thus requires attention in the future.    


\subsection{Electron Acceleration in the Buneman Zone} \label{ele-acc-buneman}

\begin{figure}[!b]
\centering
\includegraphics[width=0.98\linewidth]{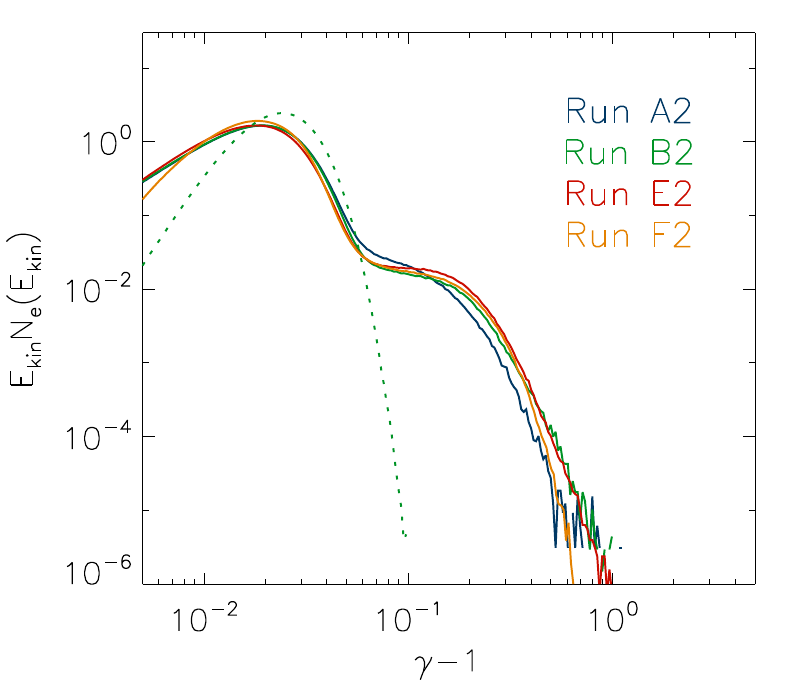}
\caption{Simulation-frame kinetic-energy spectra of electrons in the regions of the shock foot selected for Fig.~\ref{Eest_plots} color-coded for run A2 (blue), run B2 (green), run E2 (red) and for run F2 (orange). The dotted green line indicates the spectrum of upstream cold plasma electrons (extracted from run B2).}
\label{spectra_Buneman1}
\end{figure}

\begin{figure}[!b]
\centering
\includegraphics[width=0.98\linewidth]{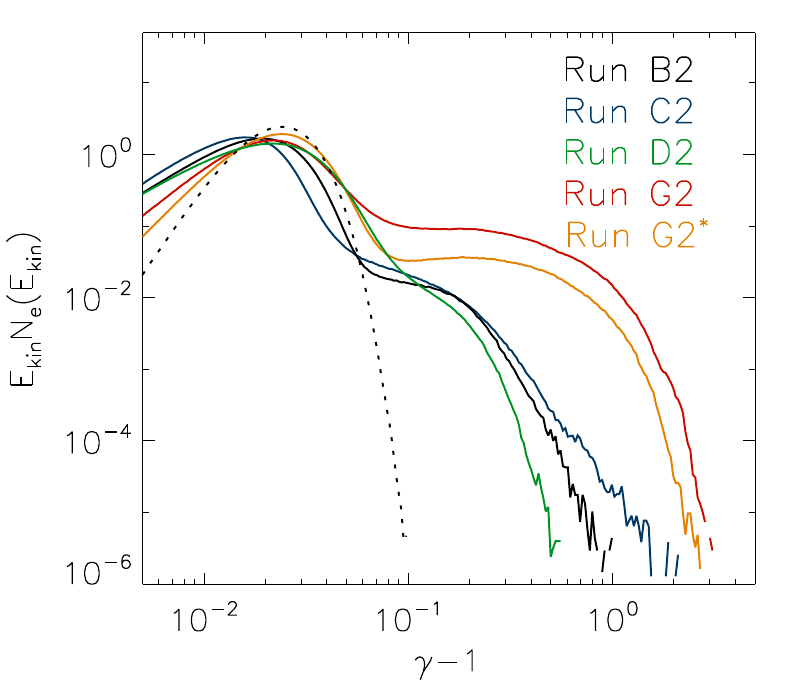}
\caption{Spectra of electrons as in Fig.~\ref{spectra_Buneman1} for run B2 (black), run C2 (blue), run D2 (green), run G2 
(red) and run G2* (orange). The dotted black line indicates the spectrum of upstream cold plasma electrons (extracted from run B2).}
\label{spectra_Buneman2}
\end{figure}

\begin{figure*}[htb!]
\centering
\includegraphics[width=0.49\linewidth]{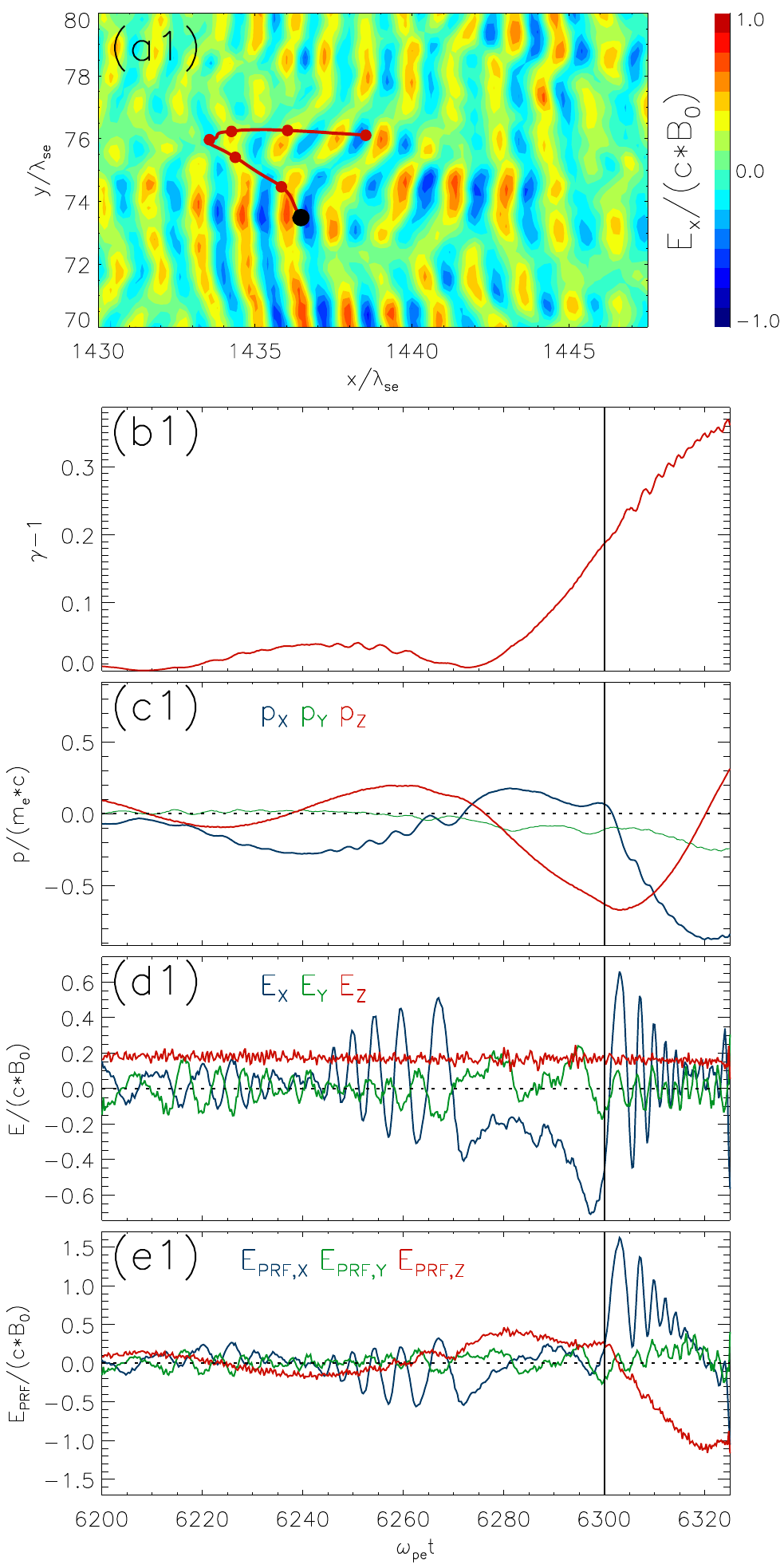}
\includegraphics[width=0.49\linewidth]{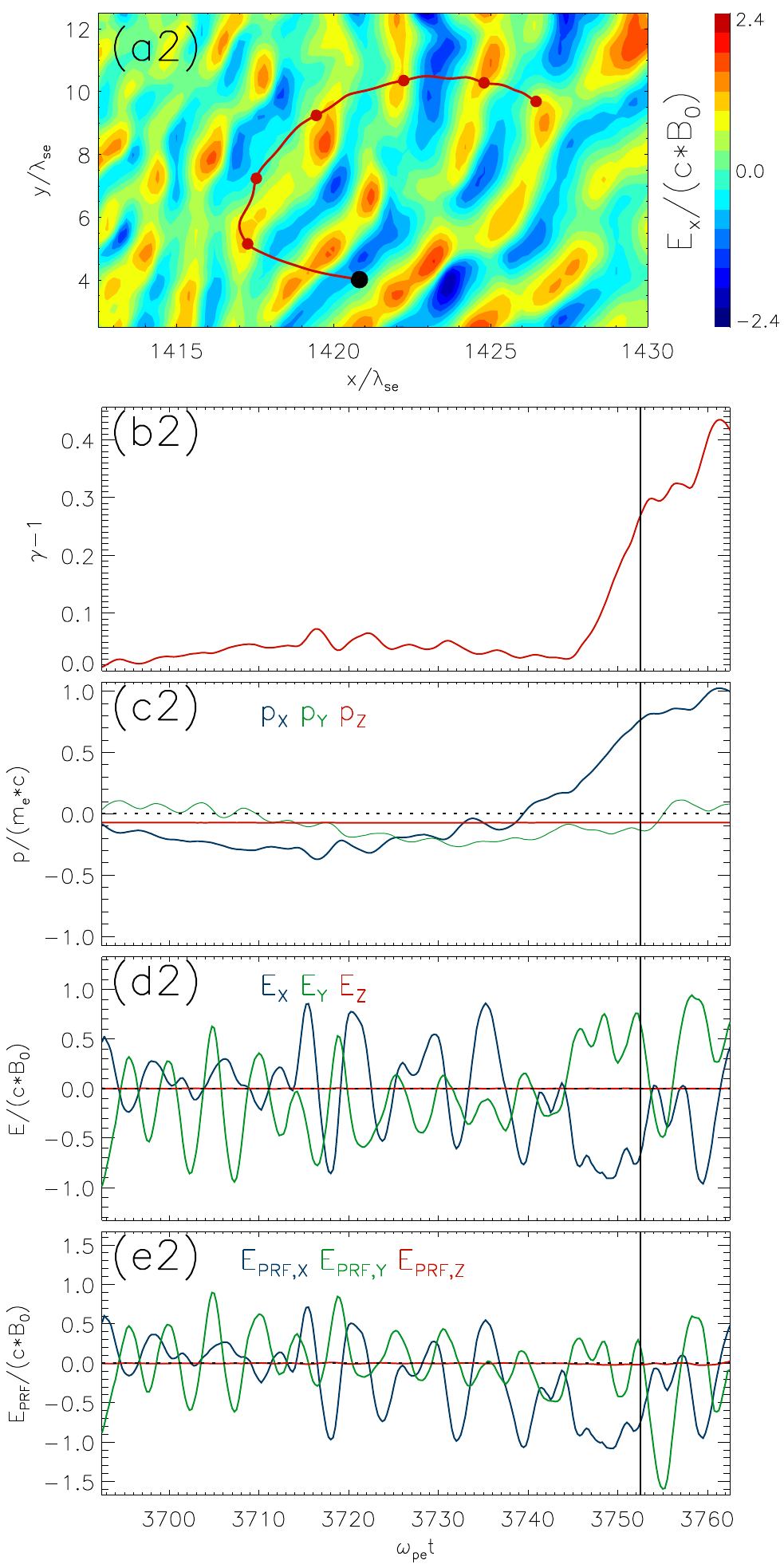}
\caption{Interaction of electrons with Buneman waves for in-plane runs (panels (a1)-(e1), case E2) and out-of-plane runs (panels (a2)-(e2), case G2). Panels (a*): map of $E_x$ at the time indicated by the vertical black lines in the lower panels. Overlaid are the position of an electron (black \rev{dot}) \revrev{at the same time moment as $E_x$ maps}, its trajectory history for the past $60\omega_{ce}^{-1}$ and \rev{past positions of the electron for every $\omega_{pe}t = 10$ intervals, designated with red dots}. Panels (b*): evolution of electron energy. Panels (c*): evolution of electron momentum. Panels (d*): dimensionless components of electric field at electron position in the simulation frame. Panels (e*): components of electric field at electron position in the electron rest frame. }
\label{acceleration-0-90}
\end{figure*}

Table~\ref{table-Buneman} lists the fraction of electrons that have been pre-accelerated in the Buneman wave zone to $(\gamma-1)>0.1$, $N_{\rm e,BI}/N_{\rm e,tot}$. This fraction is much larger in run G2 than it is in runs A2-F2. \cite{2017ApJ...847...71B} argued that at least part of this difference is due to differences in the amplitude of the electrostatic waves and their coverage area.

Figures~\ref{spectra_Buneman1} and~\ref{spectra_Buneman2} show kinetic-energy spectra of electrons occupying the Buneman wave regions highlighted in Figure~\ref{Eest_plots}.
Figure~\ref{spectra_Buneman1} shows energy spectra for runs A2, B2, E2 and F2, for which the Alfv\'enic Mach numbers exceed by the \rev{similar} margin the trapping condition (Eq.~\ref{trapping}). The spectra are statistically indistinguishable, and the fraction of pre-accelerated electrons is $\sim0.45\%$ for all runs. This is again in line with the ion-to-electron mass ratio dependence of the trapping condition.

Electron spectra for runs B2, C2, and D2, that probe different physical conditions at shocks with in-plane magnetic-field configuration, are compared to the spectrum for the out-of-plane case G2 in Figure~\ref{spectra_Buneman2}. The fractions of pre-accelerated electrons differ between the in-plane runs (see Table~\ref{table-Buneman}), reflecting the different intensities of the Buneman waves. In run C2, the spectrum extends to higher energies and contains more energetic electrons than that for run B2, which arises from the difference in Mach number. The Alfv\'enic Mach number of the shock in run D2 instead violates the trapping condition, and only $\sim0.34\%$ of electrons are pre-accelerated.  

Although run C2 satisfies the modified trapping condition of Equation~\ref{trappingnew}, the acceleration efficiency, $N_{\rm e,BI}/N_{\rm e,tot}\simeq 0.6$, is much less than for run G2. In Figure~\ref{spectra_Buneman2} the spectrum for run C2 is also compared with the spectrum calculated for run G2 at a different phase of the shock-reformation (denoted as run G2*), at which the strength of the Buneman waves matches that for run C2. Still, the fraction of pre-accelerated electrons in run G2* is four times that in run C2, but the maximum energies of the electrons are comparable, $\max(\gamma)\approx 3-4$. It is clear that the Buneman wave strength is not the only parameter that determines the efficiency of SSA in the shock foot.

SSA consists of two individual processes: (1) interaction with electrostatic waves and (2) magnetic gyration.
In the appendix we present a detailed analytical treatment of the equation of motion of electrons in the wave field, demonstrating that the electrostatic field of the waves does the physical work. Here we summarize the conclusions.

Figure~\ref{acceleration-0-90} illustrates the first stage of the SSA process for the in-plane (\emph{left} panels a1-e1) and the out-of-plane case (\emph{right} panels a2-e2). For specific electrons extracted from runs E2 and G2, we see the time evolution of the energy (Fig.~\ref{acceleration-0-90}b) and the momentum (Fig.~\ref{acceleration-0-90}c), as well as the electric field at the location of the particle in the simulation frame and in the instantaneous particle rest frame (Fig.~\ref{acceleration-0-90}d and e, respectively). The latter is particularly interesting, because in the electron rest frame the electric field is the sole provider of acceleration.
We refer to the selected electron in the in-plane case (left panels) as the first electron and the other one as the second electron. Initially both electrons move with the plasma bulk. To be trapped by electrostatic waves, the electrons must travel with the waves against the upstream plasma flow, and hence be picked-up from the thermal pool. Before doing so, the electrons move in the negative x-direction undisturbed through several electrostatic wavefronts. Significant energy gain commences at time $t\ompe=6270$ for the first electron and at 
$t\ompe=3745$ for the second electron. The particles then remain trapped by the waves and undergo the first stage of acceleration at time intervals $t\ompe=(6275-6300)$ for the first electron and $t\ompe=(3745-3753)$ for the second electron. During this stage both electrons move in the direction of shock propagation, and their $p_y$ momentum remains small. The end of the first-stage acceleration is marked by the black vertical line in Figure~\ref{acceleration-0-90}, beyond which the electrons resume gyrating.

The acceleration of the first electron occurs in the same way as in 1D geometry \citep{2002ApJ...572..880H}: the electron is pushed toward the upstream region by the electrostatic field of a Buneman wave, which for some time compensates the $x$-component of the Larmor acceleration and thus keeps the electron roughly in phase with the wave. Consequently the average values of $E_x$ and $E_y$ electric field components are close to zero in the particle reference frame (Fig.~\ref{acceleration-0-90}e1). The continuous gradient in $p_z$ at $t\ompe=(6275-6300)$ reflects the transverse Larmor acceleration, which can be described as the effect of the motional electric field in the frame of the electrostatic wave. It is important to note that for the in-plane magnetic field the wave fronts are infinitely extended in $z$-direction, and the energy gain terminates when the electron loses phase coherence with the Buneman wave. In reality the energization will terminate earlier. 
In the upstream flow frame all the energy gain comes from the field of the Buneman wave though.

The second electron displays a similar behaviour, but in the frame of the obliquely propagating waves. At $t\omega_{pe}\simeq 3745$ it starts moving in the x-direction, but the electrostatic field of the waves roughly compensates the Larmor acceleration in y-direction, as $E_{\mathrm{PRF},y} \approx 0$ (Fig.~\ref{acceleration-0-90}e2). Instead, the electron is accelerated in x-direction by the electrostatic field of the Buneman waves. 
We conclude that in all cases the energy gain arises from the electrostatic field of the waves, while the formal acceleration reflects the competition of Larmor acceleration and that imposed by the waves. The in-plane configuration captures only part of the Buneman waves, as only wave vectors in the simulation plane are allowed, and so there is a lower rate of energy gain compared to the out-of-plane case. In addition, the restriction of the wave phase velocity to the simulation plane changes the direction of sliding along a wave front from effectively the $x$-direction to the $z$-direction.

Let us estimate the energy gain arising from trapping at an electrostatic wavefront. Equations~\ref{app:e3a} and \ref{app:e7} give the rate of energy gain for the out-of-plane and the in-plane configuration, respectively. The phase speed that the electrons need to match is $v_{ph,0}=0.1c$ and $v_{ph,90}=0.4c$ for in-plane and out-of-plane configuration, respectively, and so $\Delta v$ and hence the energization rate is twice larger in the out-of-plane case than it is for in-plane magnetic field. The total energy gain is the product of the rate of gain and the time of interaction. The time of interaction is limited by three factors: the intermittency of waves, escape by acceleration perpendicular to the wave front, and escape to the side of the wave front.

In the in-plane case the wave front is infinitely extended in $z$ direction, and no escape to that side is possible. For an out-of-plane magnetic field and an average speed along the wave front of $\sim (0.1-0.2)\,c$, the electrons would escape trapping on $t_{\mathrm{esc}} \approx (25 - 50) \ompe^{-1}$, as the wave fronts in Figure~\ref{acceleration-0-90} have a lateral extent of about~$5\,\lse$.

The escape time perpendicular to the wave front can be estimated as $t_{\mathrm{esc}} \approx \pi \ompe^{-1} (v_\Phi+v_0)/v_\mathrm{e,WRF}$, where $v_\mathrm{e,WRF}$ is the velocity of electrons in the wave frame. For the out-of-plane case this gives $t_{\mathrm{esc},90} \gtrsim 18 \ompe^{-1}$, as the average electron speed $v_\mathrm{e,WRF}\lesssim 0.1\,c$.

\begin{figure}[!t]
\centering
\includegraphics[width=0.98\linewidth]{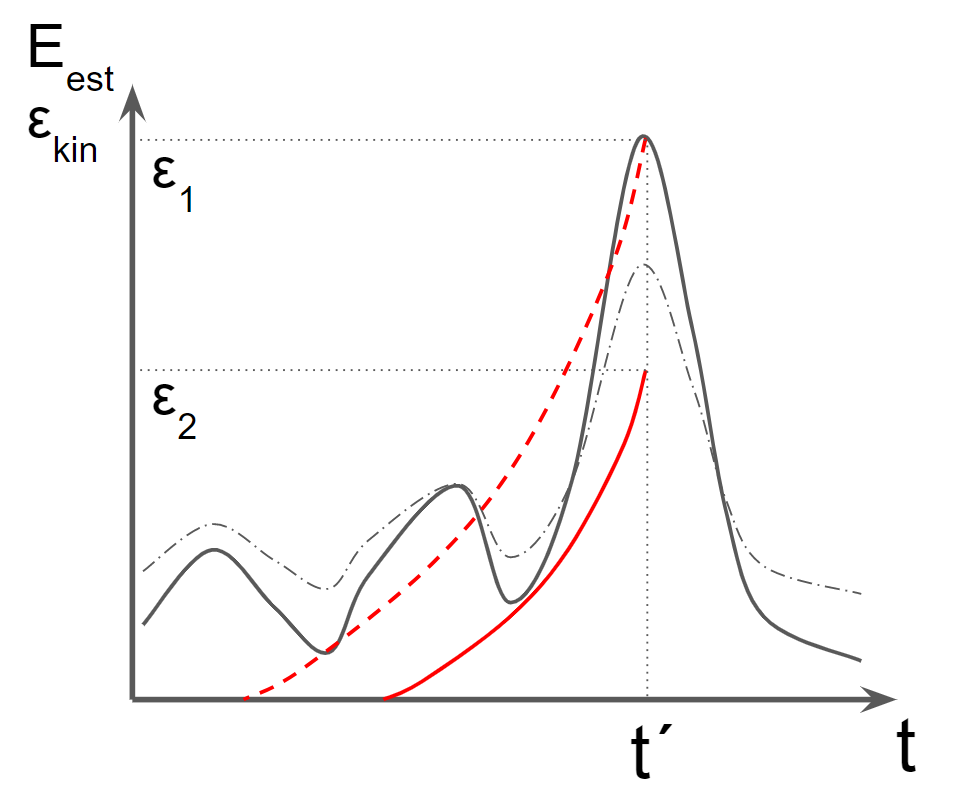}
\caption{Schematic time evolution of the electrostatic field strength (black dash-dotted line) at a chosen location in the Buneman wave rest frame. The black solid line is the maximal energy of electron can be trapped by the electrostatic field. Red lines represent energy histories  of electrons, for which trapping is possible (red solid line) and impossible (red dashed line).}
\label{trapping_time}
\end{figure}

The trapping time coming from the wave time intermittency can be estimated directly from simulations. The ability to accelerate an electron up to a certain energy depends not only on the instantaneous local electrostatic field strength but also on the previous strength history and the ability to trap an electron during the whole acceleration period. 
In Figure~\ref{trapping_time} the time evolution of electrostatic field strength ($E_\mathrm{ES}$, black dash-dotted line) at a chosen location is presented. This field is able to trap electrons with energies shown with black solid line, which is calculated assuming the equality between electrostatic and Lorentz forces at the chosen location. At a time $t'$ the electrostatic field is capable to trap an electron with energy $\epsilon_1$. However, taking into account the evolution of $E_\mathrm{ES}$ and the  energy of electrons (red dashed line) this electron cannot be trapped during the whole acceleration period. Therefore electrons with a final energy $\epsilon_2$ and the energy history shown with the red solid line can be present in the simulation.
According to these considerations the trapping time reads
\begin{equation}
t_{\mathrm{tr},0} \approx  13 \ompe^{-1} \quad\mathrm{and}\ \ 
t_{\mathrm{tr},90} \approx  11 \ompe^{-1} ,
\label{eq:tr}
\end{equation}
which is approximately the time that we analytically estimated based on the acceleration in the direction of the wave motion (see the Appendix). Thus one of the main limiting factors for electron acceleration is the intermittency of the Buneman waves.

Calculated average energy gains are
\begin{equation}
\Delta \gamma_0 \approx  0.18 \quad\mathrm{and}\ \ 
\Delta \gamma_{90} \approx  0.42 ,
\label{eq:ene}
\end{equation}
which are similar to those for electrons in Figure~\ref{acceleration-0-90} and average energies of accelerated electrons in Figures~\ref{spectra_Buneman1} and~\ref{spectra_Buneman2}. The analytically expected energy increase can be written as 
\begin{equation}
\begin{split}
\Delta \varepsilon \approx e\,E_\mathrm{ES} \vert F\vert \Delta v \,t_\mathrm{tr}= \\
= \Delta v^2 \,t_\mathrm{tr}\ompe m_e (m_e/m_i)^{(1/6)} ,
\label{eq:ene_gain}
\end{split}
\end{equation}
where $e\,E_\mathrm{ES}= m_e \Delta v\ompe (m_e/m_i)^{(1/6)}$ 
\citep{1980PhRvL..44.1404I,2009PhPl...16j2901A,2012ApJ...755..109M} and  $\vert F\vert$ is assumed to be about 1. Therefore the main difference in the acceleration rate comes from velocity difference, $\Delta v = (v_\Phi+v_0)$, and the energy gain of electrons is still stronger in the out-of-plane case due to a larger phase speed of the Buneman waves. 

\rev{We note that the modified trapping conditions (Eq.~\ref{trappingnew} or~\ref{trappingnew2}) refer to reaching a certain strength of the electrostatic field that is needed for trapping, while the energy gain of electrons is related to the velocity difference between reflected ions and upcoming electrons. This velocity difference imposes the main restriction for the in-plane simulations in their applicability to mimic realistic SSA efficiency. Using a higher Mach number can not significantly change the SSA efficiency in case of the same velocity difference defined by the magnetic field configuration. For the same mass ratio the number of pre-accelerated electrons is larger by about (30-40)\% in the runs with a higher Alfv\'enic Mach number (see Table~\ref{table-Buneman}, runs B2-C2 and D2-E2), which is not the factor of 10 required to reach the SSA efficiency seen in out-of-plain runs. Therefore significant modifications of the parameters
of the simulation (not just a change of the Alfv\'enic Mach number) are needed to reproduce the out-of-plain SSA efficiency by means of in-plane simulations.}

The energy difference associated with climbing or sliding down the potential well of a Buneman wave can be estimated as
$\Delta \gamma mc^2 = e E_{BI} \, \lambda_{BI}/2\pi$. 
The wavelength of Buneman waves, $\lambda_{BI}= 2 \pi  \Delta v /\ompe$, then implies an energy change $\Delta \gamma \approx 0.05 $ in the in-plane case and $\Delta \gamma \approx 0.17 $ in the out-of-plane run. This is insufficient to redirect an incoming electron to stationarity in the wave frame. Fluctuations in the Buneman wave field are clearly needed to trap particles and keep them in phase with the waves. 
 
We observe that for $\varphi=90\degree$ a larger number of electrons are picked up from the bulk plasma for further acceleration than is seen with the in-plane configuration, which can be explained by a twice stronger $e\,E_{\rm ES}$ force in the out-of-plane case.


\section{Summary and discussion} \label{summary}

We analyse electron injection processes at nonrelativistic perpendicular collisionless shocks with high Alfv\'enic Mach numbers with 2D3V numerical PIC simulations. Earlier studies indicated that SSA operates at the leading edge of the foot as first-stage electron pre-acceleration mechanism, provided the Alfv\'enic Mach number satisfies a condition of efficient driving of the electrostatic Buneman waves \citep[the trapping condition,][]{2012ApJ...755..109M}. In \cite{2015Sci...347..974M} and \cite{2017ApJ...847...71B} we showed that in 2D simulations that use a field component which lies in the simulation plane, the downstream nonthermal-electron fraction is much lower than with out-of-plane mean field. Noting that much of this difference results from an incomplete account of the Buneman instability in the in-plane geometry, and motivated by results of recent 3D studies which demonstrate that the injection physics past the SSA stage can adequately be studied with 2D in-plane simulations \citep{Matsumoto2017}, here we further investigate electron acceleration by SSA at perpendicular high-$\ma$ shocks with in-plane magnetic field configurations. The aim is to infer the SSA efficiency, in particular the validity of the trapping condition in its original form and the variant proposed in \cite{2017ApJ...847...71B}, and the relation to the SSA efficiency observed in simulations with the out-of-plane fields.

Our results can be summarized as follows:
\begin{itemize}

\item The energy gain in SSA always arises from the electrostatic field of a Buneman wave with which the electron travels for some time. The apparent acceleration, $\dot {\mathbf{v}}$, 
reflects the superposition of electrostatic acceleration and Larmor acceleration that might be described as effect of the motional electric field in the wave frame. This process is more efficient in the out-of-plane case because both the phase speed and the amplitude of the waves are higher than for $\varphi=0\degree$. 

\item As in high-$\ma$ shock simulations with out-of-plane magnetic fields, for in-plane magnetic field the strength of the electrostatic wave modes in the shock foot is determined by the Alfv\'enic Mach number in relation to the trapping condition. The more $\ma$ exceeds the trapping condition, the stronger the intensity of the Buneman waves. Shocks with Alfv\'enic Mach numbers satisfying the trapping condition by the \revrev{similar margin} show \revrev{comparable} wave strengths in simulations for different ion-to-electron mass ratios. 

\item Shocks in simulations with in-plane magnetic field demonstrate electrostatic wave intensities lower than those observed in the out-of-plane case, even if the modified trapping condition is satisfied. 

\item The trapping time is mostly defined by intermittency of, and limited phase-coherence of electrons with, the Buneman waves. This limits the duration of the velocity match between electrons and the waves.

\item The number of electrons pre-accelerated via SSA in the shock foot strongly correlates with the strength of the electrostatic waves. Shocks with the same physical conditions defined through the trapping condition show similar SSA efficiency. The latter is proportional to $\ma$ for a given mass ratio. However, SSA always produces larger fractions of pre-accelerated electrons in simulations with the out-of-plane configurations, even if the intensities of the Buneman waves are similar as in the in-plane case. One reason for that is the larger number of electrons being picked up from the bulk plasma for SSA compared to the in-plane configuration.    

\end{itemize}

We conclude that with an in-plane magnetic-field configuration we can not achieve the same level of SSA efficiency as in simulations with out-of-plane magnetic field or 3D simulations \citep{Matsumoto2017}, unless the parameters and settings of the simulation \rev{setup} are significantly modified.

This paper is conceived as the first of a series investigating different aspects of electron acceleration processes at non-relativistic perpendicular shocks using PIC simulations. Interaction with Weibel filaments and magnetic reconnection in the shock transition, plasma heating, and the generation of turbulent magnetic field will be covered in forthcoming publications.


\acknowledgments
\rev{We thank the anonymous referee for their comments.}
The work of J.N. has been supported by Narodowe Centrum Nauki through research project DEC-2013/10/E/ST9/00662. \rev{This work was supported by JSPS-PAN Bilateral Joint Research Project Grant Number 180500000671}.
The numerical experiment was possible through a 10 Mcore-hour allocation on the 2.399 PFlop Prometheus system at ACC Cyfronet AGH. Part of the numerical work was conducted on resources provided by the North-German Supercomputing Alliance (HLRN) under projects bbp00003 and bbp00014.


\appendix

\section{Analytical model of electron SSA}
\subsection{Out-of-plane configuration, $\varphi=90\degree$}
In the simulation frame, the large-scale magnetic field of the right plasma slab, 
$\mathbf{B}=B_0\,\hat{\boldsymbol{z}}$,
induces a motional electric field, $\mathbf{E}=-v_0\,B_0\,\hat{\boldsymbol{y}}$, where $v_0$ is the speed of the upstream plasma flowing in $-x$ direction. The entire Larmor orbit of all particles with low temperature is leveled in the simulation plane, as are the acceleration imposed by the waves. 

Suppose an electrostatic wave propagates at an angle $\Theta$ to the x-axis. The electric field carried by the wave is 
\begin{equation}
E_x=E_\mathrm{ES}\,F\,\cos\Theta\qquad E_y=E_\mathrm{ES}\,F\,\sin\Theta,
\label{app:e1}
\end{equation}
where the wave factor is
\begin{equation}
F=\sin\left(\frac{\ompe}{v_\Phi+v_0}\left[x\,\cos\Theta+y\,\sin\Theta-v_\Phi t\right]+\Phi\right).
\label{app:e2}
\end{equation}
Here we allow for an arbitrary phase, $\Phi$. The phase speed of the wave, $v_\Phi$, is measured in the simulation frame. The wave number is related to the velocity of reflected ions through the resonance condition of the Buneman modes, $\ompe=k (v_\Phi+v_0)$. 

Now consider an electron with velocity components $v_x$ and $v_y$.
Using non-relativistic kinematics we find the acceleration of the electron as
\begin{align}
\dot v_x=&-\omce \frac{E_\mathrm{ES}}{B_0} F \cos\Theta-\omce v_y \nonumber\\
\dot v_y=&\omce v_0 +\omce v_x -\omce \frac{E_\mathrm{ES}}{B_0} F \sin\Theta.
\label{app:e3}
\end{align}
Let us rotate the coordinate system by an angle $\Theta$, so that $x'$ is oriented in the direction of motion of the waves and $y'$ is perpendicular to it. The corresponding accelerations then read
\begin{align}
\dot v_{x'}=&\omce\left(v_0 \sin\Theta - \frac{E_\mathrm{ES}}{B_0} F - v_{y'}\right) \nonumber\\
\dot v_{y'}=&\omce \left(v_0 \cos\Theta + v_{x'} \right).
\label{app:e4}
\end{align}

The wave factor, $F$, is explicitly time-dependent and may induce rapidly oscillating acceleration. The other terms only describe Larmor gyration in the flow frame and hence no real energy gain. The wave factor must be approximately constant, if continuous energy gain is to be achieved for about 10 plasma times, $\ompe^{-1}$, as observed. This requires that on average $v_{x'}-v_\Phi \lesssim 0.2\,c$ or roughly acceleration from $v_{x'}=0.2\,c$ to $v_{x'}=0.6\,c$, after which the electron is out of phase with the wave and commences Larmor motion.

The Larmor motion of the reflected ions mandates a wave direction for which $\sin\Theta$ is negative. Likewise, the wave factor, $F$, must be negative to effect energy gain. Equation~\ref{app:e4} then indicates that acceleration in $y'$ direction follows that in $x'$ direction, and for a fair range of initial conditions $\dot v_{y'}$ is slightly less than $\dot v_{x'}$ and increases with the same rate, at least for up to $1\,\omce^{-1}\simeq 12\,\ompe^{-1}$. Correspondingly, the momentum component $p_x$ increases approximately linearly, and the increase in speed is approximately $E_\mathrm{ES}/(2B_0)$, whereas $p_y$ remains approximately constant.

The effective acceleration toward the upstream region arises from the superposition of acceleration in the electrostatic field of the Buneman waves and the Larmor acceleration, that are oppositely directed in $y$ direction, but both have positive components in $x$ direction. Acknowledging that both $F$ and $\sin\Theta$ must be negative,
the rate of energy gain in the upstream flow frame is
\begin{equation}
m\frac{d}{dt}\frac{(v_x+v_0)^2+v_y^2}{2}=eE_\mathrm{ES}\vert F\vert
\left[(v_x+v_0)\cos\Theta -v_y\vert \sin\Theta\vert\right]
\label{app:e3a}
\end{equation}
and hence completely independent of the motional electric field. In the simulation frame the velocity component $v_0 \cos\Theta$ disappears from Equation~\ref{app:e3a} and a new component of energy-gain rate appears, $\omce v_0 v_y$, which captures the apparent energy by Larmor motion in this frame.

\subsection{In-plane configuration, $\varphi=0\degree$}
The main impact of the in-plane configuration is that the part of the Larmor motion is perpendicular to the simulation plane, and so the orientation and properties of the Buneman waves are modified, as only wave vectors in the simulation plane can be captured. The wave factor changes to 
\begin{equation}
F=\sin\left(\frac{\ompe}{v_\Phi+v_0}\left[x-v_\Phi t\right]+\Phi\right).
\label{app:e5}
\end{equation}
The acceleration then follows by appropriate rotation of that given in Equation~\ref{app:e3}, 
\begin{align}
\dot v_x=&-\omce \frac{E_\mathrm{ES}}{B_0} F +\omce v_z \nonumber\\
\dot v_z=&-\omce v_0 -\omce v_x .
\label{app:e6}
\end{align}
Obviously, there is linear acceleration in $-z$ direction, if the particle can be held at approximately constant phase ($F< 0$; $v_x \approx v_\Phi$) in the wave. As $v_\Phi \gtrsim v_0$ it is the Larmor acceleration that is responsible for the particle's sliding along the wavefront, and the electrostatic field of the waves provides slow energy gain at a rate
\begin{equation}
m\frac{d}{dt}\frac{(v_x+v_0)^2+v_z^2}{2}=eE_\mathrm{ES}\vert F\vert
(v_x+v_0) ,
\label{app:e7}
\end{equation}
which also only involves the electrostatic field of the Buneman waves. The energy gain will be less than that for out-of-plane configuration, because only part of the motion of the back-streaming ions can drive waves that hence have lower amplitude, $E_\mathrm{ES}$, and additionally the velocity term in Equation~\ref{app:e7} is reduced.




\end{document}